\documentclass[aps,preprint,amsmath,amssymb,12pt]{revtex4}
\usepackage{graphicx,color}

\def\be{\begin{eqnarray}}
\def\ed{\end{eqnarray}}
\def\non{\nonumber}
\def\hmp{\hat m_{P}}
\def\hml{\hat m_{\ell}}

\def\pp{{\prime\prime}}

\begin{document}

\preprint{BARI-TH-09-617}

\title{\Large \bf $|V_{ub}|$ and  $B\to\eta^{(')}$ Form Factors  in  Covariant  Light-Front Approach  }

\author{ \bf  Chuan-Hung Chen$^{1,2}$\footnote{Email:
physchen@mail.ncku.edu.tw}, Yue-Long Shen$^{3}$\footnote{Email:
shenyl@phys.sinica.edu.tw} and Wei Wang$^{4}$\footnote{Email:
wei.wang@ba.infn.it}
 }

\affiliation{ $^{1}$Department of Physics, National Cheng-Kung
University, Tainan 701, Taiwan \\
$^{2}$National Center for Theoretical Sciences, Taiwan\\
$^{3}$Institute of Physics, Academia Sinica, 115 Taipei, Taiwan \\
$^{4}$  Istituto Nazionale di Fisica Nucleare, Sezione di Bari, Bari
70126, Italy
 }

\date{\today}

\begin{abstract}
$B\to (\pi, \eta, \eta')$ transition form factors are investigated
in the covariant light-front approach. With theoretical
uncertainties, we find that $B\to (\pi, \eta, \eta')$ form factors
at $q^2=0$ are $f^{(\pi, \eta,
\eta')}_{+}(0)=\left(0.245^{+0.000}_{-0.001}\pm 0.011,\, 0.220 \pm
0.009\pm0.009,\, 0.180\pm 0.008^{+0.008}_{-0.007}\right)$ for vector
current and $f^{(\pi, \eta, \eta')}_{T}(0)=\left(
0.239^{+0.002+0.020}_{-0.003-0.018},\, 0.211\pm
0.009^{+0.017}_{-0.015},\, 0.173\pm 0.007^{+0.014}_{-0.013}\right)$
for tensor current, respectively. With the obtained $q^2$-dependent
$f^{\pi}_{+}(q^2)$ and observed branching ratio (BR) for $\bar
B_d\to \pi^+ \ell \bar \nu_{\ell}$, the $V_{ub}$ is found as
$|V_{ub}|_{LF}= (3.99 \pm 0.13)\times 10^{-3}$. As a result, the
predicted BRs for $\bar B\to (\eta, \eta') \ell \bar\nu_{\ell}$
decays with $\ell=e,\mu$ are given by
$\left(0.49^{+0.02+0.10}_{-0.04-0.07},\,
0.24^{+0.01+0.04}_{-0.02-0.03}\right)\times 10^{-4}$, {while the BRs
for $D^-\to (\eta,\eta')\ell\bar\nu_{\ell}$ are
$(11.1^{+0.5+0.9}_{-0.6-0.9},\,
1.79^{+0.07+0.12}_{-0.08-0.12})\times 10^{-4}$. In addition, we also
study the integrated lepton angular asymmetries for $\bar B\to
(\pi,\eta,\eta')\tau \bar\nu_{\tau}$:
$(0.277^{+0.001+0.005}_{-0.001-0.007},0.290^{+0.002+0.003}_{-0.000-0.003},0.312^{+0.004+0.005}_{-0.000-0.006})$.}

\end{abstract}

\maketitle %

One of the puzzles in exclusive $B$ decays is the $\eta'$ related
processes. For example  $B \to \eta^{\prime} K$ was first observed
by CLEO collaboration with the branching ratio (BR)
$(89^{+18}_{-16}\pm 9)\times 10^{-6}$ \cite{CLEO_PRL81} which is
much larger than $30 \times 10^{-6}$ estimated by the factorization
ansatz. With more data accumulated, experimental uncertainties are
pinning down and the world averages on the BR for $\eta^{(\prime)}$
production in $B$ decays now are known as ${\cal B}(B^{+}\to [\eta,
\eta'] K^{+})=[ 2.36\pm0.27, 71.1\pm 2.6]\times 10^{-6}$, ${\cal
B}(B^{0}\to [\eta, \eta'] K^{0})=[ 1.12^{+0.30}_{-0.28}, 66.1\pm
3.1]\times 10^{-6}$ \cite{HFAG}. Clearly, the large BR for $B\to
\eta' K$ is not smeared by statistic. To unravel the mystery, many
novel solutions have been proposed, such as the intrinsic charm in
$\eta^{\prime}$ \cite{icharm}, the gluonium state \cite{gluon}, the
spectator hard scattering mechanism \cite{2yang}, the flavor-singlet
component of $\eta^{\prime}$ \cite{BN_NPB} and enhanced chiral
symmetry breaking effects \cite{ACG}. Although it is believed that
some exotic effects should be associated with $\eta^{(\prime)}$, it
is difficult to specify where the novel effects should reside, since
two-body hadronic $B$ decays  suffer from large uncertainties such
as final state interactions.

Compared with nonleptonic $B$ decays,  semileptonic $\bar B\to
\eta^{(\prime)} \ell \bar\nu_{\ell}$ and $\bar B_s\to
\eta^{(\prime)} \ell^+\ell^-$ decays  are much cleaner and thus
might be more helpful to explore the differences among various
mechanisms. In particular, a sizable flavor-singlet component of
$\eta^{(\prime)}$ predicts larger BRs for $\bar B\to\eta' \ell
\bar\nu_{\ell}$ than the $\eta$ modes, while the chiral symmetry
breaking enhancement could give the reverse
results \cite{ACG}. 
Nevertheless, before one considers various possible novel effects on
$\eta^{(\prime)}$, it is necessary to understand  the BRs for $\bar
B\to \eta^{(\prime)} \ell \bar\nu_{\ell}$ decays without these
exotic effects. In our previous work \cite{ACG}, we used the
perturbative QCD approach~\cite{Keum:2000ph} to calculate the $B\to
\eta^{(\prime)}$ form factors at large recoil; then the same whole
spectrum as a function of invariant mass of $\ell \nu_\ell$ for the
form factors is assumed with that in the light-cone sum rules
(LCSRs). Despite the predicted results for various branching ratios
are consistent with the experimental data, it is meaningful to
examine the same processes in other parallel frameworks. This is
helpful to reduce the dependence on the treatments of the dynamics
in transition form factors.  The motif of this work is to employ
another method to deal with the form factors: the covariant
light-front (LF) approach \cite{Jaus:1999zv,CCH}.  Since the
predictions of $B\to \pi$ form factors in LF model match very well
with those applied to the nonleptonic charmless $B$ decays, it is
worthy to understand what we can get the $B\to \eta^{(\prime)}$ form
factors by this approach.

%

At the quark level, the $\bar B\to  \eta^{(\prime)} \ell
\bar\nu_{\ell}$ is induced by $b\to ul\bar\nu$ transition  which
will inevitably involve the $\bar uu$ component of the $\eta^{(')}$
meson. Then the convenient mechanism for the $\eta-\eta'$ mixing
would be the quark flavor mixing scheme, defined by
\cite{flavor0,flavor}
\begin{eqnarray}
\left( {\begin{array}{*{20}c}
   \eta   \\
   {\eta '}  \\
\end{array}} \right) = \left( {\begin{array}{*{20}c}
   {\cos \phi } & { - \sin \phi }  \\
   {\sin \phi } & {\cos \phi }  \\
\end{array}} \right)\left( {\begin{array}{*{20}c}
   {\eta _{q} }  \\
   {\eta _{s} }  \\
\end{array}} \right) \,,\label{eq:flavor}
\end{eqnarray}
where $\eta _{q}  = ( {u\bar u + d\bar d})/\sqrt{2}$, $\eta_{s} =
s\bar s $ and angle $\phi$ is the mixing angle. By the definition of
$\langle 0| \bar q' \gamma_{\mu} \gamma_{5} q'|
\eta_{q'}(p)\rangle=if_{q'}p_{\mu}$ ($q'=q,s$), the masses of
$\eta_{q,s}$ can be expressed by
\begin{eqnarray}
m_{qq}^2  &=& \frac{\sqrt 2}{f_{q} }\langle 0|m_u \bar ui\gamma _5 u
+ m_d \bar di\gamma _5 d| \eta_q \rangle,\ \ \ m_{ss}^2  =
\frac{2}{f_{s}}\langle 0 |m_s \bar si\gamma _5 s| \eta _s \rangle.
\label{eq:masses}
\end{eqnarray}
Here, $m_{qq}$ and $m_{ss}$ are unknown parameters and their values
can be obtained by fitting with the data. In terms of the
quark-flavor basis, we see clearly that $m_{qq}$ and $m_{ss}$ are
zero in the chiral limit. The advantage of the quark-flavor mixing
scheme is: at the leading order in $\alpha_s$ only the quark
transition from the $B$ meson into the $\eta_q$ component is
necessary; while the other transitions like $B\to\eta_s$ are
suppressed by $\alpha_s$. The gluonic form factors (or referred to as
flavor-singlet form factors) will be remarked later.

For calculating the transition form factors, we parameterize the
hadronic effects as
\begin{eqnarray}
\langle P(P'') | \bar q^{\prime} \gamma^{\mu}  b| \bar B(P')\rangle
&=& f^{P}_{+}(q^2)\left(P^{\mu}-\frac{P\cdot q}{q^2}q^{\mu}
\right)+f^{P}_{0}(q^2) \frac{P\cdot q}{q^2} q_{\mu} \,, \non
\\
\langle P(P'' )| \bar q^{\prime} i\sigma_{\mu\nu} q^{\nu}b| \bar B
(P')\rangle &=& {f^{P}_{T}(q^2)\over m_{B}+m_{P}}\left[P\cdot q\,
q_{\mu}-q^{2}P_{\mu}\right]\,,\non \\
\langle P(P'' )| \bar q^{\prime} \sigma_{\mu\nu} \gamma_5 b| \bar B
(P')\rangle &=& \frac{f^{P}_{T}(q^2)}{
m_{B}+m_{P}}\epsilon_{\mu\nu\alpha\beta}P^\alpha q^\beta
 \label{eq:bpff}
\end{eqnarray}
with $P_{\mu}=(P'+P'')_{\mu}$ and  $q_{\mu}=(P'-P'')_{\mu}$. Since
the light quarks in $B$-meson are $u$- and $d$-quark, the meson $P$
could stand for $\pi$ and $\eta_q$ states.

\begin{figure}
\includegraphics[scale=0.5]{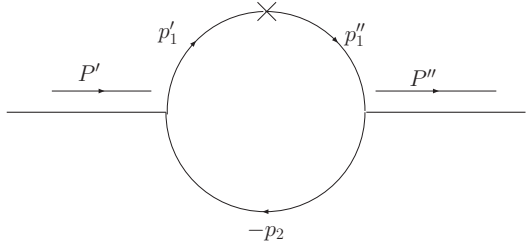}
\caption{Feynman diagram for the transition form factors, where the
cross symbol in the diagram denotes the  transition
vertex.}\label{fig:feyn}
\end{figure}
In the covariant LF quark model,
the transition form factors for $B\to P$ could be obtained by
computing the lowest-order Feynman diagram depicted in
Fig.\ref{fig:feyn}.  Below we will adopt the same notation as
Ref.~\cite{Jaus:1999zv} and light-cone coordinate system for
involved momenta, in which the components of meson momentum are
read by $P^{\prime}=(P^{\prime -}, P^{\prime +}, P^\prime_\bot)$
with $P^{\prime\pm}=P^{\prime0}\pm P^{\prime3}$. The relationship
between meson momentum and the momenta of its constitutent quarks
is given by $P^{\prime}=p_1^{\prime}+p_2$ and
$P^{\pp}=p_1^{\pp}+p_2$ with
$p_2$ being the spectator quark of initial and final
mesons. Additionally, one can also express the quark momenta in
terms of the internal variables $(x_i, p_\bot^\prime)$ as
 \begin{eqnarray}
 \tilde p_{1,2}^{+}=x_{1,2} \tilde P^{ +},\qquad
 \tilde p_{1,2\bot}=x_{1,2} \tilde P_\bot\pm \tilde p_\bot
 \end{eqnarray}
with $x_1+x_2=1$. Here, the notation with tilde could represent all
momenta in the initial and final mesons.

In order to formulate the results of Fig.~\ref{fig:feyn}, the
quark-meson-antiquark vertex for incoming and outgoing mesons are
respectively chosen to be
\begin{eqnarray}
&&i\Gamma_P^\prime = H^\prime_P\gamma_5\,,\non \\
&& i(\gamma_0 \Gamma^{\prime\dagger}_P\gamma_0)\,,
\end{eqnarray}
where $H^{\prime}_{P}$ is the covariant light-front wave function of
the meson. Consequently, the amplitude for the loop diagram is
straightforwardly written by
\begin{eqnarray}
 {\cal B}^{PP}_{\mu(\mu\nu)}=-i^3\frac{N_c}{(2\pi)^4}\int d^4 p^\prime_1
 \frac{H^\prime_P ( H^\pp_P)}{N_1^\prime N_1^\pp N_2} S^{PP}_{\mu(\mu\nu)},
\end{eqnarray}
where $N_c=3$ is the number of colors, $N_1^{\prime(\prime\prime)}=
p_1^{\prime(\prime\prime)2} -m_1^{\prime (\prime\prime)2} $,
$N_2=p_2^2-m_2^2$.
\begin{eqnarray}
 S^{PP}_\mu   &=&2 p^\prime_{1\mu} [M^{\prime 2}+M^{\pp2}-q^2-2
           N_2-(m_1^\prime-m_2)^2-(m^\pp_1-m_2)^2+(m_1^\prime-m_1^\pp)^2]
 \nonumber\\
          &&+q_\mu[q^2-2 M^{\prime2}+N^\prime_1-N^\pp_1+2
          N_2+2(m_1^\prime-m_2)^2-(m_1^\prime-m_1^\pp)^2]
 \nonumber\\
          &&+P_\mu[q^2-N^\prime_1-N^\pp_1-(m_1^\prime-m_1^\pp)^2]\,,
          \nonumber \\S^{PP}_{\mu\nu }  &=&-\epsilon_{\mu\nu\alpha\beta}(-4m_2p_1^{'\alpha} q^\beta+2m_1''p_1^{'\alpha}q^{\beta}-2m_1'' P^\alpha p_1^{\prime\beta}
 \non\\
 &+&2m_1' P^\alpha p_1^{\prime\beta}-2m_1'P^\alpha q^\beta+2m_1'p_1^{\prime\alpha}
 q_\beta)\,,
 \label{eq:SPPVappen}
\end{eqnarray}
with the $M'(M'')$ being the mass of the incoming (outgoing) meson.
As usual, the loop integral could be performed by the contour
method. Therefore, except some separate poles appearing in the
denominator, if the covariant vertex functions are not singular, the
integrand is analytic. Thus, when performing the integration, the
transition amplitude will pick up the singularities from the
anti-quark propagator so that the various pieces of integrand are
led to be
 \be
 N_1^{\prime(\pp)}
      &\to&\hat N_1^{\prime(\pp)}=x_1(M^{\prime(\pp)2}-M_0^{\prime(\pp)2}),
 \nonumber\\
 H^{\prime(\pp)}_P
      &\to& h^{\prime(\pp)}_P,
 \nonumber\\
 S
      &\to& \hat S,
 \nonumber\\
\int \frac{d^4p_1^\prime}{N^\prime_1 N^\pp_1 N_2}H^\prime_P H^\pp_P
S
      &\to& -i \pi \int \frac{d x_2 d^2p^\prime_\bot}
                             {x_2\hat N^\prime_1
                             \hat N^\pp_1} h^\prime_P h^\pp_P \hat
                             S.
 \ed
We work in the $q^+=0$ frame and the transverse momentum of the
quark in the final meson is given as
$p^\pp_\bot=p^\prime_\bot-x_2\,q_\bot$. The new function of
$h^\prime_M$ for initial meson is given by
\begin{eqnarray}
 h^\prime_P&=&(M^{\prime2}-M_0^{\prime2})\sqrt{\frac{x_1 x_2}{N_c}}
                    \frac{1}{\sqrt{2}\widetilde M^\prime_0}\varphi^\prime_P\,,
\label{eqs:wa}
\end{eqnarray}
with
\begin{eqnarray}
 M^{\prime2}_0
          &=&(e^\prime_1+e_2)^2=\frac{p^{\prime2}_\bot+m_1^{\prime2}}
                {x_1}+\frac{p^{\prime2}_{\bot}+m_2^2}{x_2}\,,\quad\quad
                \widetilde
                M^\prime_0=\sqrt{M_0^{\prime2}-(m^\prime_1-m_2)^2}\,,
 \nonumber\\
 e^{(\prime)}_i
          &=&\sqrt{m^{(\prime)2}_i+p^{\prime2}_\bot+p^{\prime2}_z}\,,\quad\qquad
 p^\prime_z=\frac{x_2 M^\prime_0}{2}-\frac{m_2^2+p^{\prime2}_\bot}{2 x_2
 M^\prime_0}\,,
 \end{eqnarray}
where $e_i$ can be interpreted as the energy of the quark or the
antiquark, $M_0^\prime$ can be regarded as the kinetic invariant
mass of the meson system and $\varphi'_P$ is the LF momentum
distribution amplitude for $s$-wave pseudoscalar mesons. The similar
quantities associated with the outgoing meson can be defined by the
same way.

After the contour integration, the valance antiquark is turned to be
on mass-shell and the conventional LF model is recovered. The
formulas of the form factors in the LF quark model shown in
Eq.~(\ref{eq:SPPVappen}) would contain not only the terms
proportional to $P_\mu$ and $q_\mu$, but also the terms proportional
to a null vector $\tilde\omega=(2,0,{\bf 0_\perp})$. This vector is
spurious, because it does not appear in the standard definition of
Eq.(\ref{eq:bpff}), and spoils the covariance. In the literature, it
is argued that this spurious factor can be eliminated by including
the so-called zero-mode contribution, and a proper way to resolve
this problem has been proposed in Ref.~\cite{Jaus:1999zv}. In this
method, one should obey a series of special rules when performing
the $p^-$ integration. A manifest covariant result can be given with
this approach, which is physically reasonable. Using
Eqs.~(\ref{eq:SPPVappen})--(\ref{eqs:wa}) and taking the advantage
of the rules in Ref.~\cite{Jaus:1999zv,CCH}, the $B\to P$ form
factors are straightforwardly obtained by
  \be
 f^{P}_+(q^2)&=&\frac{N_c}{16\pi^3}\int dx_2 d^2p^\prime_\bot
            \frac{h^\prime_P h^\pp_P}{x_2 \hat N_1^\prime \hat N^\pp_1}
            \bigg[x_1 (M_0^{\prime2}+M_0^{\pp2})+x_2 q^2
 \non\\
         &&\qquad-x_2(m_1^\prime-m_1^\pp)^2 -x_1(m_1^\prime-m_2)^2-x_1(m_1^\pp-m_2)^2\bigg]\,,
 \nonumber\\
 f^{P}_-(q^2)&=&\frac{N_c}{16\pi^3}\int dx_2 d^2p^\prime_\bot
            \frac{2h^\prime_P h^\pp_P}{x_2 \hat N_1^\prime \hat
            N^\pp_1}
            \Bigg\{- x_1 x_2 M^{\prime2}-p_\bot^{\prime2}-m_1^\prime m_2
                  +(m_1^\pp-m_2)(x_2 m_1^\prime+x_1 m_2)
\non\\
         &&\qquad +2\frac{q\cdot P}{q^2}\left(p^{\prime2}_\bot+2\frac{(p^\prime_\bot\cdot q_\bot)^2}{q^2}\right)
                  +2\frac{(p^\prime_\bot\cdot q_\bot)^2}{q^2}
                  -\frac{p^\prime_\bot\cdot q_\bot}{q^2}
                  \Big[M^{\pp2}-x_2(q^2+q\cdot P)
\non\\
         &&\qquad -(x_2-x_1) M^{\prime2}+2 x_1 M_0^{\prime
                  2}-2(m_1^\prime-m_2)(m_1^\prime+m_1^\pp)\Big]
           \Bigg\}\,,\non
           \\
  f^{P}_T(q^2)&=&
           { (M'+M'')}\frac{N_c}{16\pi^3}\int dx_2
d^2p^\prime_\bot
            \frac{h^\prime_P h^\pp_P}{x_2 \hat N_1^\prime \hat N^\pp_1}
            \bigg[x_1(2m_2-m_1'-m_1'')+2m_1' \non \\
            &-&2(m_1'-m_1'')\left(\frac{x_1}{2}-\frac{p^\prime_\bot\cdot q_\bot}{q^2}\right)\bigg]\,,
 \label{eq:fpm}
 \ed
where the relation of $f^{P}_{-}(q^2)$ to $f^{P}_{0}(q^2)$ can be
read by
 \be
 f^{P}_0(q^2)=f^{P}_+(q^2)+\frac{q^2}{m_{B}^2-m_P^2}f^{P}_-(q^2).
\label{eq:a4}
 \ed
Clearly, one has $f^{P}_{+}(0)=f^{P}_{0}(0)$.

After we obtain the formulae for the $B\to P$ transition form
factors, the direct application is the exclusive semileptonic $\bar
B\to P \ell \bar\nu_{\ell}$ decays. The effective Hamiltonian for
$b\to u \ell \bar{\nu}_{\ell}$ in the standard model (SM) is given
by
\begin{eqnarray}
H_{\rm eff}&=& \frac{G_FV_{ub}}{\sqrt{2}}  \bar u\gamma_{\mu}
(1-\gamma_5) b\, \bar \ell \gamma^{\mu} (1-\gamma_5) \nu_{\ell}\,.
 \label{eq:heff}
\end{eqnarray}
Although these decays are tree processes, however, if we can
understand well the form factors, there still have the chance to
probe the new physics in these semileptonic decays \cite{CG,CN}.
Hence, the decay amplitude for $\bar B\to P \ell \bar \nu_{\ell}$ is
written as
 \be
M(\bar B\to P \ell \bar \nu_{\ell})&=&\langle  \ell \bar\nu_{\ell}
P| H_{\rm eff}|\bar B\rangle = \frac{G_F V_{ub}}{\sqrt{2}} \langle P
| \bar u \gamma_{\mu}(1-\gamma_5)b| \bar B\rangle \bar
\ell\gamma^{\mu} (1-\gamma_5) \nu_{\ell}\,.
    \label{eq:amp_p}
 \ed
To calculate the differential decay rates, we choose the coordinates
of various particles as follows
\begin{eqnarray}
q^{2}&=&(\sqrt{q^2},\,0,\, 0,\, 0), \ \ \ p_{B}=(E_{B},\, 0,\, 0,\, |{\cal \bf p}_{P}|), \nonumber \\
p_{P}&=& (E_{P},\, 0,\, 0,\, |{\cal \bf p}_P|), \ \ \
p_{\ell}=(E_{\ell},\, |{\cal \bf p}_{\ell}| \sin\theta,\, 0,\,
|{\cal \bf p}_{\ell}| \cos\theta)\,, \label{eq:coordinates}
\end{eqnarray}
where $E_{P}=(m^{2}_{B}-q^2-m^2_{P})/(2\sqrt{q^2})$, $|{\cal \bf
p}_{P}|=\sqrt{E^2_{P}-m^2_{P}}$,
$E_{\ell}=(q^2+m^2_{\ell})/(2\sqrt{q^2})$ and $|{\cal \bf
p}_{\ell}|=(q^2-m^2_{\ell})/(2\sqrt{q^2})$. It is clear that
$\theta$ is defined as the polar angle of the lepton momentum
relative to the moving direction of the $B$-meson in the $q^2$ rest
frame. With Eqs.~(\ref{eq:amp_p}) and (\ref{eq:coordinates}), the
differential decay rate for $\bar B\to P \ell \bar\nu_{\ell}$ as a
function of $q^2$ and $\theta$ can be derived by
\begin{eqnarray}
\frac{d\Gamma_P}{dq^2 d\cos\theta}&=& \frac{G^{2}_{F} |V_{ub}|^2
m^3_{B}}{2^8
\pi^3}\sqrt{(1-s+\hmp^2)^2-4\hmp^2}\left(1-\frac{\hml^2}{s}
\right)^2 \non \\
&& \times \left[ \Gamma^{P}_{1}+\Gamma^{P}_{2} \cos\theta +
\Gamma^{P}_{3} \cos^2\theta \right]\,, \label{eq:diff_P}
\\
\Gamma_1^P&=& \hat P_P^2f_+^{P2}(q^2)
+(1-\hat m_P^2)^2 \frac{\hat m_{\ell}^2}{s} f_0^{P2}(q^2),\nonumber\\
\Gamma_2^P&=&2 \frac{\hat m_{\ell}^2}{s} \hat P_P (1-\hat m_P^2)
f^{P}_+(q^2)f^{P}_0(q^2),\nonumber\\
\Gamma_3^P&=& -\hat P_P^2 f_+^{P2}(q^2)+ \frac{\hat m_{\ell}^2}{s}
\hat P_P^2 f_+^{P2}(q^2),\non
 \label{eq:gamma_P}
\end{eqnarray}
where $s=q^{2}/m_{B}^{2}$, $\hat m_{i}=m_{i}/m_{B}$ and
 \begin{eqnarray}
 \hat P_{P}&=&2\sqrt{s} |{\cal
\bf p}_{P}|/m_{B}=\sqrt{(1-s-\hmp^2 )^2-4s\hmp^2}\,.\label{eq:hatPP}
\end{eqnarray}
Since the differential decay rate in Eq.~(\ref{eq:diff_P}) involves
the polar angle of the lepton, we can define an angular asymmetry to
be
\begin{eqnarray}
{\cal A}(q^2)={\int^{1}_{-1}dz {\rm sign}(z) d\Gamma_P/(dq^2dz)
\over \int^{1}_{-1}dz d\Gamma_P /(dq^2dz)}\label{eq:asy}
\end{eqnarray}
with $z=\cos\theta$. Explicitly, the asymmetry for $\bar B\to P \ell
\bar \nu_{\ell}$ decay is
\begin{eqnarray}
{\cal A}_{P}(s)={\Gamma^{P}_{2} \over 2 \Gamma^{P}_{1}+ 2/3
\Gamma^{P}_{3}}\,. \label{eq:asy_P}
\end{eqnarray}
Moreover, the integrated angular asymmetry can be defined by
 \be
{\cal \bar A}_{P} &=& { \int dq^2 \int^{1}_{-1}dz {\rm sign}(z)
d\Gamma_P/(dq^2dz) \over \int dq^2 \int^{1}_{-1}dz d\Gamma_P
/(dq^2dz)}\,. \label{eq:int_asy}
 \ed
The angular asymmetry is only associated with the ratio of form
factors, which supposedly is insensitive to the  hadronic
parameters. Plausibly, this physical quantity could be the good
candidate to explore the new physics such as charged Higgs
\cite{CG}, right-handed gauge boson \cite{CN}, etc.

Before presenting the numerical results for the form factors and
other related quantities, we will briefly discuss how to extract the
input parameters for the $\eta_q$ in the presence of  $\eta-\eta'$
mixing. Following the divergences of the axial vector currents
\begin{eqnarray}
\partial ^\mu  \bar q^{\prime} \gamma_{\mu}\gamma_5 q^{\prime}
  &=& \frac{ \alpha _s }{4\pi }G
\tilde G +  2 m_{q^{\prime}} \bar{q}^{\prime}i\gamma _5 q^{\prime}
,
\label{eq:axial}
\end{eqnarray}
where $G=G^{a\mu\nu}$ are the gluonic field-strength and $\tilde
G=\tilde G^{a\mu\nu}\equiv\epsilon^{\mu\nu\alpha
\beta}G^{a}_{\alpha\beta}$,  the mass matrix of $\eta_{q,s}$ becomes
\begin{eqnarray}
\left( {\begin{array}{*{20}c}
   {M_{qq}^2 } & {M_{qs}^2 }  \\
   {M_{sq}^2 } & {M_{ss}^2 }  \\
\end{array}} \right) & = &\left( {\begin{array}{*{20}c}
   \langle 0|\partial^\mu  J_{\mu 5}^q | {\eta _q }\rangle /f_q  & \langle 0 |\partial^\mu  J_{\mu 5}^s | \eta _q \rangle /f_s  \\
   \langle 0|\partial^\mu  J_{\mu 5}^q | {\eta _s } \rangle /f_q & \langle 0 |\partial^\mu  J_{\mu 5}^s | \eta _s  \rangle/f_s   \\
\end{array}} \right) \nonumber \\
  & = & \left( {\begin{array}{*{20}c}
   m_{qq}^2  + 2a^2 & \sqrt{2} y a^2  \\
   \sqrt{2} y a^2 & m_{ss}^2  + y^2 a^2       \\
\end{array}} \right)
\end{eqnarray}
with $ a^2 = \langle 0| \alpha_s G \tilde
G|\eta_q\rangle/(4\sqrt{2}\pi f_{q})$ and $y=f_{q}/f_{s}$. Using the
mixing matrix introduced in Eq.~(\ref{eq:flavor}), one can
diagonalize the mass matrix and  the eigenvalues are the physical
mass of $\eta$ and $\eta'$. Correspondingly, we have  the relations
\cite{FK}
\begin{eqnarray}
\sin\phi &=& \left[\frac{(m^{2}_{\eta^{\prime}}-m^2_{ss})
(m^2_{\eta}-m^2_{qq})}{(m^{2}_{\eta^{\prime}}-m^2_{\eta})(m^2_{ss}-m^2_{qq})}
\right]^{1/2}\,, \non \\
y &=& \left[2\frac{(m^{2}_{\eta^{\prime}}-m^2_{ss})
(m^2_{ss}-m^2_{\eta})}{(m^{2}_{\eta^{\prime}}-m^2_{qq})(m^2_{\eta}-m^2_{qq})}
\right]^{1/2}\,, \non \\
a^2&=&\frac{1}{2} \frac{(m^{2}_{\eta^{\prime}}-m^2_{qq})
(m^2_{\eta}-m^2_{qq})}{m^2_{ss}-m^2_{qq}}\,, \label{eq:parameters}
 \end{eqnarray}
and $m_{\eta^{(\prime)}}$ is the mass of $\eta^{(\prime)}$. Once the
parameters $\phi$, $y$ and $a$ are determined by experiments, we can
get the information for $m_{qq,ss}$ and $f_{q,s}$. Then, they could
be taken as the inputs in our calculations.

After formulating the necessary pieces, we now perform the numerical
analysis for the form factors and the related physical quantities
introduced earlier. For understanding how well the predictions of LF
model are, we first analyze $B\to \pi$ form factors at $q^2=0$. By
examining Eq.~(\ref{eq:fpm}), we see that the main theoretical
unknowns are the parameters of distribution amplitudes of mesons,
masses of constitute quarks and the decay constants of mesons. As
usual, we adopt the gaussian-type wave function for pseudoscalar
mesons as
 \be
%
     \varphi_P^\prime(x_2,p^\prime_\perp)
           &=&4 \left({\pi\over{\beta_P^{\prime2}}}\right)^{3\over{4}}
               \sqrt{{dp^\prime_z\over{dx_2}}}~{\rm exp}
               \left(-{p^{\prime2}_z+p^{\prime2}_\bot\over{2 \beta_P^{\prime2}}}\right),
\label{eq:Gauss}
 \ed
with $\beta'_{P}$ characterizing the shape of the wave function.
Other relevant values of parameters are taken as (in units of GeV)
 \begin{eqnarray}
m_{B}&=&5.28  \,,\ \ \  m_b= (4.8\pm0.2) \,, \ \ \ m_{\pi}=0.14 \,, \non\\
m_{u}&=&m_{d}=(0.26\pm0.03)  \,, \ \ \ f_{B}=(0.19\pm0.02)  \,, \ \
\
f_{\pi}=0.131 \,,\non\\
\beta'_B &=& 0.553^{+0.047}_{-0.048}\,, \ \ \
\beta'_{\pi}=0.31,\,\;\;\;\;\;
\beta'_{\eta_q}=0.353_{-0.013}^{+0.014}\,,\label{eq:inputparameters}
 \end{eqnarray}
where $m_{u,d}$ are the constituent quark masses,  the errors in
them are from the combination of linear, harmonic oscillator and
power law potential \cite{Choi} and $f_{P}$ denotes the decay
constant of P-meson. The shape parameters $\beta s$ are determined by
the relevant decay constants whose analytic expressions are given in
Ref.~\cite{CCH}. Following the formulae derived in
Eq.~(\ref{eq:fpm}) and using the taken values of parameters, we
immediately find
 \be
f^{\pi}_{+}(0)&= &0.245^{+0.000+0.011}_{-0.001-0.011}\,, \ \ \
f^{\pi}_{T}(0)= 0.239^{+0.002+0.020}_{-0.003-0.018}\,,
  \ed
where the two kinds of uncertainties are from (i) $\beta_B'$; (ii)
the quark masses $m_u$ and $m_b$ (added in quadrature). To compare
with the results of LCSRs given by $f^\pi_{+}(0)|_{LCSR}=0.258\pm
0.031$ and $f^{\pi}_{T}(0)|_{LCSR}= 0.253\pm 0.028$ \cite{LCSR}, it
is clear that although the central value of LF model is slightly
smaller than those of LCSRs, they are still consistent with each
other by counting the errors.   Since we use the quark-flavor
scheme, for estimating the form factors associated with
$\eta^{(\prime)}$,
the values of involving parameters are chosen to be $f_{q}=(1.07\pm
0.02) f_{\pi}$, $\phi=39.3^{\circ}\pm 1.0^{\circ}$\cite{flavor},
$m_{qq}=0.14^{+0.11}_{-0.04}$ GeV \cite{ACG},
$f^{\eta}_{+(T)}=\cos\phi f^{\eta_q}_{+(T)}$ and
$f^{\eta'}_{+(T)}=\sin\phi f^{\eta_q}_{+(T)}$, we have
 \be
 f^{\eta_q}_{+}(0)&=& 0.284^{+0.012+0.012}_{-0.012-0.011}\,
 \,, \ \ \ f^{\eta_q}_{T}(0)=
 0.273^{+0.011+0.022}_{-0.011-0.020}\,, \non
  \\
f^{\eta}_{+}(0)&=& 0.220^{+0.009+0.009}_{-0.009-0.009}\,
 \,, \ \ \ f^{\eta}_{T}(0)=
 0.211^{+0.009+0.017}_{-0.009-0.015}\,, \non
  \\
 f^{\eta'}_{+}(0)&=& 0.180^{+0.008+0.008}_{-0.008-0.007}\,
 \,, \ \ \ f^{\eta'}_{T}(0)=
 0.173^{+0.007+0.014}_{-0.007-0.013}\,,\label{eq:formfactors-eta-zero-point}
 \ed
where the first and second errors are from  (i) $\beta'_B$ and
$\beta'_{\eta_q}$ (ii) the quark masses $m_u$ and $m_b$,
respectively. {  From Eq.~\eqref{eq:fpm}, one can see that the
form factor $f^{\eta_q}_+(q^2)$ does not depend on the mass
$m_{qq}$, while the dependence of $m_{qq}$ in $f^{\eta_q}_T$
resides in the term $M'+M''$ (in this case $m_B+m_{qq}$). The
uncertainty of $f_T$ caused by the $m_{qq}$ is less than $2\%$.}
Furthermore, since the form factors are associated with mixing
angle $\phi$, the corresponding uncertainties for $B\to
\eta^{(\prime)}$ and BRs of $\bar B\to \eta^{(\prime)} \ell \bar
\nu_{\ell}$ are expected to be $2.1\%$ ($1.4\%$) and $4.2\%$
($2.8\%$), respectively. Despite different treatments of quarks'
momenta, the results here are well consistent with that in
light-cone quark model constructed in the effective field
theory~\cite{Lu:2007sg}:
$f_+^{\eta_q}(0)=0.287^{+0.059}_{-0.065}$. Intriguingly, our
results are also consistent with $f^{\eta}_{+}(0)|_{LCSR}=
0.231^{+0.018}_{- 0.020}$ and
$f^{\eta'}_{+}(0)|_{LCSR}=0.189^{+0.015}_{-0.016}$ calculated by
LCSRs \cite{BJ}. In order to understand the behavior of whole
$q^2$, the form factors for $B\to P$ are parametrized by
\cite{LCSR}
 \be
 F_i(q^2)&=&\frac{F_i(0)}{1-aq^2/m_B^2+b(q^2/m_B^2)^2}\,, \label{eq:ff}
 \ed
where $F_i$ denotes any form factor among $f_{+,0,T}$. The fitted
values of $a,b$ for $B\to (\pi,\, \eta,\, \eta')$ are displayed in
Table~\ref{tab:LFFF}, where the uncertainties are similar to the
ones given in Eq.~(\ref{eq:formfactors-eta-zero-point}).
\begin{table}[httb]
\caption{ Values of parameters for $q^2$ dependent $B\to (\pi, \eta,
\eta')$ form factors calculated by LF quark model. }\label{tab:LFFF}
\begin{ruledtabular}
\begin{tabular}{cccccc}
 $F(q^2)$  & $a$ & $b$  & $F(q^2)$ & $a$ & $b$
 \\ \hline

 $f^{B\to \pi}_{+}$ & $1.62^{+0.05+0.12}_{-0.05-0.11}$ & $0.79^{+0.09+0.17}_{-0.08-0.15}$ & $f^{B\to \eta^{(\prime)}}_{+}$ & $1.55^{+0.04+0.11}_{-0.05-0.10}$ & $0.65^{+0.08+0.13}_{-0.06-0.12}$ \\
 $f^{B\to \pi}_{0}$ & $0.75^{+0.04+0.11}_{-0.03-0.10}$ & $0.07^{+0.02+0.05}_{-0.03-0.05}$ & $f^{B\to \eta^{(\prime)}}_{0}$ & $0.67^{+0.03+0.09}_{-0.03-0.09}$ & $0.03^{+0.03+0.03}_{-0.02-0.03}$ \\
 $f^{B\to \pi}_{T}$ & $1.60^{+0.05+0.12}_{-0.05-0.11}$ & $0.75^{+0.09+0.17}_{-0.08-0.15}$ &$f^{B\to \eta^{(\prime)}}_{T}$ & $1.53^{+0.04+0.11}_{-0.04-0.10}$ & $0.62^{+0.08+0.13}_{-0.07-0.11}$
\\
\end{tabular}
\end{ruledtabular}
\end{table}

In the quark flavor mixing mechanism, the $\eta$ and $\eta'$ meson
receives additional coupling with two gluons, due to the axial
anomaly. Thus to be self-consistent, in the study of the transition
form factors, one also needs to include the so-called gluonic form
factors which is induced by the transition from the two gluons into
the $\eta^{(')}$. {In our study, the gluonic form factors have been
neglected and there are two reasons for this.  In the light-front
quark model, the leading order contribution to the form factor is of
the order $\alpha_s^0$ while the gluonic form factor is suppressed
by the $\alpha_s$, where the coupling constant is evaluated at the
typical scale $\mu\sim \sqrt {\Lambda_{\rm QCD}\times m_B}$ (with
$\Lambda_{\rm QCD}$ hadronic scale). The inclusion of the gluonic
form factors also requires the next-to-leading order studies for the
quark content, which is beyond the scope of the present work.
Secondly the factorization analysis of the gluonic form factors such
as the perturbative QCD study in Ref.~\cite{Charng:2006zj} reflects
that there is no endpoint singularity in the gluonic form factors
and the PQCD study shows that the gluonic form factors are
negligibly small. This feature is also confirmed by the recent LCSR
results~\cite{BJ}. For terms without endpoint singularity, different
approaches usually obtain similar results. Thus our results of the
semileptonic $B\to \eta^{(\prime)}l\bar\nu$ will not be sizably
affected by the gluonic form factors, although they are not taken
into account in the present analysis.}

Besides the form factors could be the source of uncertainties,
another uncertain quantity in exclusive $b\to u \ell \bar
\nu_{\ell}$ decays is from the Cabibbo-Kobayashi-Maskawa (CKM)
matrix element $V_{ub}\sim \lambda^3$ with $\lambda$ being
Wolfenstein parameter.  Results for $V_{ub}$ determined by inclusive
and exclusive decaying modes have some inconsistencies
\cite{PDG_08,CN}. For a self-consistent analysis, we take $B\to \pi$
form factors calculated by LF model and the data ${\cal B}(\bar
B_d\to \pi^+ \ell' \bar\nu_{\ell'})=(1.36\pm0.09)\times 10^{-4}$
with $\ell'= e,\mu$~\cite{PDG_08} as the inputs to determine the
$|V_{ub}|$. Neglecting the lepton mass, one gets the differential
decaying rate for $\bar B\to \pi \ell' \bar\nu_{\ell'}$
 \be
\frac{d\Gamma_{\pi}}{dq^2 }&=& \frac{G^{2}_{F} |V_{ub}|^2
m^3_{B}}{3\cdot 2^6
\pi^3}\sqrt{(1-s+\hat{m}_{\pi}^2)^2-4\hat{m}_{\pi}^2}
\left(f^{\pi}_{+}(q^2) \hat P_{\pi}\right)^2\,,
 \label{eq:diffpi}
 \ed
where only the $f^{\pi}_{+}$ form factor involves. Accordingly, the
value of $V_{ub}$ is found by
 \be
 |V_{ub}|_{LF}= (3.99 \pm 0.13)\times 10^{-3}\,.
 \ed
With the obtained result of $|V_{ub}|_{LF}$, the form factors in the
Table~\ref{tab:LFFF}, 
the predicted BRs for $B^-\to (\eta,\,\eta') \ell \bar\nu_{\ell}$,
together with the experimental results measured by BaBar
collaboration~\cite{Babar_eta}, are displayed in
Table~\ref{table:LFBR}.  The predicted result for the BR of $B^-\to
\eta \ell \bar\nu_{\ell}$ is about two times larger than that of
$B^-\to \,\eta'  \ell \bar\nu_{\ell}$: the form factor of $B^-\to
\eta$ is larger than the form factor of $B^-\to \eta'$; the phase
space in $B\to \eta' \ell \bar\nu_{\ell}$ is smaller. Branching
ratios for decays with a tau lepton are naturally smaller than the
relevant channels with a lighter lepton.
%
 \begin{table}[hptb]
\caption{   BRs of
 $B^{-}\to \eta^{(\prime)} \ell \bar
 \nu_{\ell}$ (in units of $10^{-4}$). The two kinds of errors shown
in the table are from (i) $\beta'_B$, $\beta'_{\eta_q}$; (ii) the
quark masses $m_u,m_b$, respectively.}   \label{table:LFBR}
\begin{ruledtabular}
\begin{tabular}{ccccccc}
Mode  & $B^{-}\to \eta \ell' \bar \nu_{\ell'}$ & $B^-\to \eta \tau
\bar \nu_{\tau}$ & $B^{-}\to \eta^{\prime} \ell' \bar \nu_{\ell'}$ &
$B^-\to \eta' \tau \bar\nu_\tau$
 \\ \hline 
 This work &$0.49^{+0.02+0.10}_{-0.04-0.07}$ &$0.29^{+0.01+0.07}_{-0.02-0.05}$
 &$0.24^{+0.01+0.04}_{-0.02-0.03}$ & $0.13^{+0.01+0.03}_{-0.01-0.02}$ \\ \hline
 Exp.~\cite{Babar_eta} & $0.31\pm 0.06\pm 0.08$ & & $<0.47$ &
 \end{tabular}
\end{ruledtabular}
\end{table}

\begin{figure}[bpth]
\includegraphics[scale=0.9]{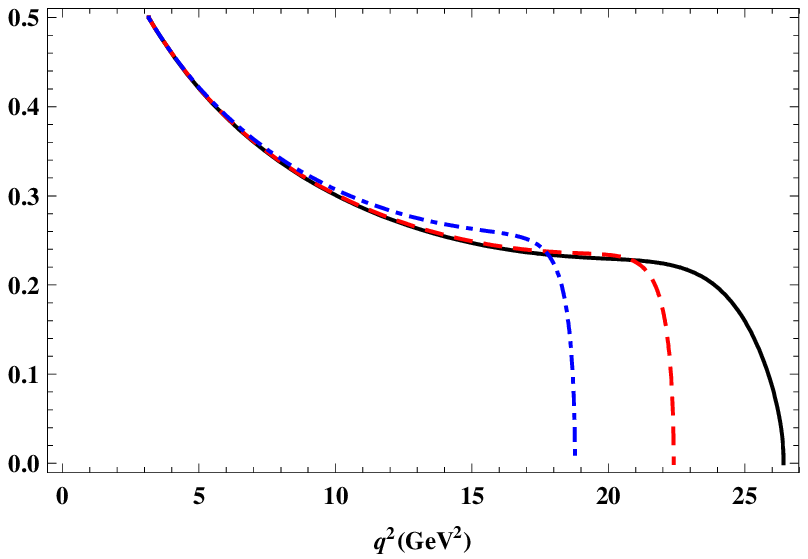}
\put (-250,100){${\cal A}_p(s)$} \caption{Angular asymmetries for
$B\to (\pi,\eta,\eta')\ell\nu_{\ell}$ are depicted by the solid
(black), dashed (red) and dash-dotted (blue) lines, respectively.
}\label{fig:angular-asymmetry}
\end{figure}

According to Eq.~(\ref{eq:asy_P}), moreover, we can study the lepton
angular asymmetries. Using the obtained form factors, we present the
asymmetry as a function of $q^2$ in
Fig.~\ref{fig:angular-asymmetry}, where the solid, dashed and
dash-dotted lines are for $B^-\to(\pi^0,\, \eta,\, \eta') \tau
\bar\nu_\tau$, respectively. Due to the angular asymmetry being
proportional to $m^2_\ell$ in the SM, here we only present the
effects on $\tau$ decaying modes. At very small $q^2$ region, the
three lines are approaching the point $0.75$ which can be easily
derived from the definition of angular asymmetries.  At the small
recoil region, the $\hat P_P$ defined in the Eq.~(\ref{eq:hatPP}) is
approaching zero and all the lepton angular asymmetries are  close
to 0. {  The integrated angular asymmetries defined in
Eq.~(\ref{eq:int_asy}) for $B^-\to(\pi^0,\, \eta,\, \eta') \tau
\bar\nu_\tau$ are predicted by
$(0.277^{+0.001+0.005}_{-0.001-0.007},0.290^{+0.002+0.003}_{-0.000-0.003},0.312^{+0.004+0.005}_{-0.000-0.006})$.}

Finally, we make some remark on the $D$ decays. We find that the
obtained information on $\eta^{(\prime)}$ can be directly applied to
the semileptonic $D^+\to \eta^{(\prime)} \ell \nu_{\ell}$ decays.
Since the associated CKM matrix element $|V_{cd}|=0.2256 \pm 0.0010$
has small errors, if the decay constant of $D$-meson is well
controlled, the $\eta^{(\prime)}$ production in $D$ decays could be
the good environment to test the properties of $\eta^{(\prime)}$.
Hence, by similar calculations performed in $B$ decays and taking
$f_{D}=(0.205\pm0.020)$ GeV, $\beta_{D}=(0.462^{+0.048}_{-0.047})$
GeV and $m_c=(1.4\pm0.1)$ GeV,  the form factors for $f^{D\to
\eta^{(\prime)}}_{+,0,T}(q^2)$ are obtained by
 \be
 f^{D\to \eta_q}_{+}(q^2) &=& \frac{0.688}{1-1.03\hat s+0.29\hat
 s^2},\nonumber\\
 f^{D\to \eta_q}_{0}(q^2) &=& \frac{0.705}{1-0.39\hat s+0.01\hat
 s^2},\nonumber\\
 f^{D\to \eta_q}_{T}(q^2) &=& \frac{0.616}{1-1.08\hat s+0.25\hat
 s^2},
 \ed
where we have used the parametrization defined in Eq.~(\ref{eq:ff}),
$\hat s=q^2/m_D^2$ and  only the central values are shown. The small
differences between $f_+(0)$ and $f_0(0)$ arise from the fitting
procedure. Replacing the parameters of $B$-meson appearing in
Eq.~(\ref{eq:diffpi}) by those of $D$-meson, the BRs for $D^-\to
(\eta,\, \eta')\ell \bar \nu_{\ell}$ are predicted by
 \be
{\cal B}(D^-\to \eta \ell \bar\nu_\ell) &=& (1.11^{+0.05+0.09}_{-0.06-0.09})\times 10^{-3}\,,\non \\
{\cal B}(D^-\to \eta' \ell \bar\nu_\ell) &=&
(1.79^{+0.07+0.12}_{-0.08-0.12})\times 10^{-4}\,,
 \ed
respectively. It is found that ${\cal B}(D^-\to \eta' \ell
\bar\nu_\ell)$ is almost one order of magnitude smaller than ${\cal
B}(D^-\to \eta \ell \bar\nu_\ell)$. The reason for the resulted
smallness is just phase space suppression. Our predictions are well
consistent with the recent measurements by the CLEO
collaboration~\cite{Mitchell:2008kb}:
\begin{eqnarray}
{\cal B}(D^-\to \eta \ell \bar\nu_\ell) &=&
(1.33\pm0.20\pm0.06)\times 10^{-3},\nonumber\\
{\cal B}(D^-\to \eta' \ell \bar\nu_\ell) &<& 3.5\times 10^{-4}.
\end{eqnarray}
This consistence is very encouraging.  The $D^-\to \eta' l\bar\nu$
may be detected in the near future. Our results are also consistent
with the results given in Ref.~\cite{Wei:2009nc}.

In summary, we have calculated the $B\to (\pi,\, \eta,\, \eta')$
transition form factors in LF approach. We find that at maximum
recoil the values of form factors are $f^{(\pi, \eta,
\eta')}_{+}(0)=\left(0.245^{+0.000}_{-0.001}\pm 0.011,\, 0.220 \pm
0.009\pm0.009,\, 0.180\pm 0.008^{+0.008}_{-0.007}\right)$ and
$f^{(\pi, \eta, \eta')}_{T}(0)=\left(
0.239^{+0.002+0.020}_{-0.003-0.018},\, 0.211\pm
0.009^{+0.017}_{-0.015},\, 0.173\pm 0.007^{+0.014}_{-0.013}\right)$,
respectively. Our calculated values are consistent with the results
done by LCSRs. With the obtained form factor $f^{\pi}_{+}(q^2)$ and
observed BR for $\bar B_d\to \pi^+ \ell \bar \nu_{\ell}$, the
$V_{ub}$ is extracted to be $|V_{ub}|_{LF}= (3.99 \pm 0.13)\times
10^{-3}$. Accordingly, we predict the BRs for $\bar B\to (\eta,
\eta') \ell \bar\nu_{\ell}$ decays with $\ell=e,\mu$  as
$\left(0.49^{+0.02+0.10}_{-0.04-0.07},\,
0.24^{+0.01+0.04}_{-0.02-0.03}\right)\times 10^{-4}$, {  while the
BRs for $D^-\to (\eta,\eta')\ell\bar\nu_{\ell}$ are given by
$(11.1^{+0.5+0.9}_{-0.6-0.9},\,
1.79^{+0.07+0.12}_{-0.08-0.12})\times 10^{-4}$. In addition, we also
show that the lepton angular asymmetries for $\bar B\to
(\pi,\eta,\eta')\tau \bar\nu_{\tau}$ are $(0.277^{+0.001+0.005}_{-0.001-0.007},0.290^{+0.002+0.003}_{-0.000-0.003},0.312^{+0.004+0.005}_{-0.000-0.006})$.} \\

{\bf Acknowledgments}\\

This work is supported by the National Science Council of R.O.C.
under Grant No: NSC-97-2112-M-006-001-MY3, and National Natural
Science Foundation of China under the Grant No. 10847161.

\end{document}